\documentclass[aps,prl,twocolumn,showpacs,preprintnumbers,amsmath,amssymb,superscriptaddress]{revtex4}%

\usepackage{graphicx}%
\usepackage{dcolumn}
\usepackage{amsmath}
\usepackage{color}
\usepackage{multirow}

\begin{document}


\title{Nodeless superconductivity in quasi-one-dimensional Nb$_2$PdS$_5$: a $\mu$SR study}

\author{P.~K.~Biswas}
\email[Corresponding author: ]{pabitra.biswas@psi.ch}
\affiliation{Laboratory for Muon Spin Spectroscopy, Paul Scherrer Institute, CH-5232 Villigen PSI, Switzerland}
\author{H.~Luetkens}
\affiliation{Laboratory for Muon Spin Spectroscopy, Paul Scherrer Institute, CH-5232 Villigen PSI, Switzerland}
\author{Xiaofeng~Xu}
\affiliation{Department of Physics and Hangzhou Key Laboratory of Quantum Matters, Hangzhou Normal University, Hangzhou 310036, China}
\author{J.~H.~Yang}
\affiliation{Department of Physics and Hangzhou Key Laboratory of Quantum Matters, Hangzhou Normal University, Hangzhou 310036, China}
\author{C.~Baines}
\affiliation{Laboratory for Muon Spin Spectroscopy, Paul Scherrer Institute, CH-5232 Villigen PSI, Switzerland}
\author{A.~Amato}
\affiliation{Laboratory for Muon Spin Spectroscopy, Paul Scherrer Institute, CH-5232 Villigen PSI, Switzerland}
\author{E.~Morenzoni}
\affiliation{Laboratory for Muon Spin Spectroscopy, Paul Scherrer Institute, CH-5232 Villigen PSI, Switzerland}
\date{\today}

\begin{abstract}
Muon spin relaxation and rotation ($\mu$SR) measurements have been performed to study the superconducting and magnetic properties of Nb$_2$PdS$_5$. Zero-field $\mu$SR data show that no sizeable spontaneous magnetization arises with the onset of superconductivity in Nb$_2$PdS$_5$ which indicates that the time reversal symmetry is probably preserved in the  superconducting state of this system. A strong diamagnetic shift is observed in the transverse-field (TF) $\mu$SR data practically ruling out a dominant triplet-pairing superconducting state in Nb$_2$PdS$_5$. The temperature dependence of magnetic penetration depth evidences the existence of a single \textit{s}-wave energy gap, $\Delta(0)$ with a gap value of 1.07(4) meV at zero temperature. The ratio $\Delta(0)/k_{\rm B}T_{\rm c}=2.02(9)$ indicates that Nb$_2$PdS$_5$ should be considered as a moderately strong-coupling superconductor. The magnetic penetration depth at zero temperature is 785(20) nm, indicating a very low superfluid density consistent with the quasi-one-dimensional nature of this system.
\end{abstract}

\maketitle


The recent discovery of superconductivity in the quasi-one-dimensional (Q1D) material Nb$_2$PdS$_5$ with a $T_{\rm c}$ of 6.6~K has attracted interest due to its remarkably large and anisotropic upper critical field ($H_{\rm c2}>370$ kOe along the \textit{b}-axis)~\cite{Zhang1,Zhang2}, far above its Pauli paramagnetic limit ($H_{\rm p}=121$~kOe). Layered superconductors generally exhibit large $H_{\rm c2}$ values, which often significantly exceed the paramagnetic limit. Examples include high-$T_{\rm c}$ cuprates~\cite{Vedeneev} and Fe-based superconductors~\cite{Altarawneh,Jaroszynski,Yuan,Tarantini}. There are several circumstances under which $H_{\rm c2}$ may exceed the Pauli limit. These are strong spin-orbit scattering~\cite{Ferrell,Anderson}, multiband effect~\cite{Gurevich1,Gurevich2}, spatially inhomogeneous Fulde-Ferrel-Larkin-Ovchinnikov (FFLO) state~\cite{Fulde,Larkin,Matsuda} and spin-triplet pairing~\cite{Ferrell,Anderson}.

Nb$_2$PdS$_5$ crystallizes in the low-dimensional monoclinic space group $C2/m$~\cite{Zhang1}. Density functional theory calculations show that the Fermi surface (FS) is composed of Quasi-2D sheets of hole character and strongly warped Quasi-1D sheets with both electron and hole characters~\cite{Zhang1}. The large $H_{\rm c2}$ in this system was tentatively ascribed to multiband effects or spin-triplet pairing~\cite{Zhang1,Zhang2}. Recently it has been shown that $H_{\rm c2}$ of this system can be modified in a tunable fashion by chemical doping on the Pd chains with elements of varying mass numbers and hence by altering the spin-orbit coupling on the Pd sites~\cite{Zhou}. It was also suggested that superconductivity in Nb$_2$PdS$_5$ may appear in proximity to a magnetic instability. Very recent specific heat measurements on this material however show that the size of the heat capacity jump at $T_{\rm c}$ is smaller than the BCS value and that a nodeless gap opens at the Fermi surface~\cite{Niu}. All these experimental data are far from conclusive and the exact nature of its unconventional SC phases has yet to be resolved. To gain further information about the superconducting properties and magnetic state of this transition metal-chalcogenide-based Q1D compound, we have performed a detailed $\mu$SR study of this system. Our zero-field (ZF)-$\mu$SR results find no evidence of any sizeable magnetism in the superconducting state of Nb$_2$PdS$_5$. Transverse-field (TF)-$\mu$SR results show that, at low temperature, the superfluid density $\rho_s\propto\lambda^{-2}$ ($\lambda$ is the magnetic penetration depth) becomes temperature independent, which is consistent with a fully gapped superconducting state. $\rho_s(T)$ can be well fitted with a single-gap $s$-wave model with a gap value of 1.07(4) meV at absolute zero temperature. This results in a gap to $T_{\rm c}$ ratio of 2.02(9) suggesting a strong coupling of the superconducting charge carriers. The absolute value of the magnetic penetration depth is determined to be $\lambda(0)=785(20)$~nm.


Polycrystalline samples of Nb$_2$PdS$_5$ were grown by a solid-state reaction method~\cite{Zhang1} and characterized as described in Ref.~\cite{Niu}. The resistivity (with a standard four-probe technique) and DC magnetization measurements were performed in a Quantum Design Physical Property Measurement System (PPMS) and Quantum Design Magnetic Property Measurement System (MPMS), respectively. The TF- and ZF-$\mu$SR experiments were carried out in the LTF instrument at the $\pi$M3 beam line of the Paul Scherrer Institute (Villigen, Switzerland). The sample was cooled to the base temperature in zero field for the ZF-$\mu$SR experiments and in 695.7~Oe for the TF-$\mu$SR experiments. Typically $\sim12$~million muon decay events were collected for each spectrum. The ZF- and TF-$\mu$SR data were analyzed by using the free software package MUSRFIT~\cite{Suter}.


\begin{figure}[htb]
\includegraphics[width=1.0\linewidth]{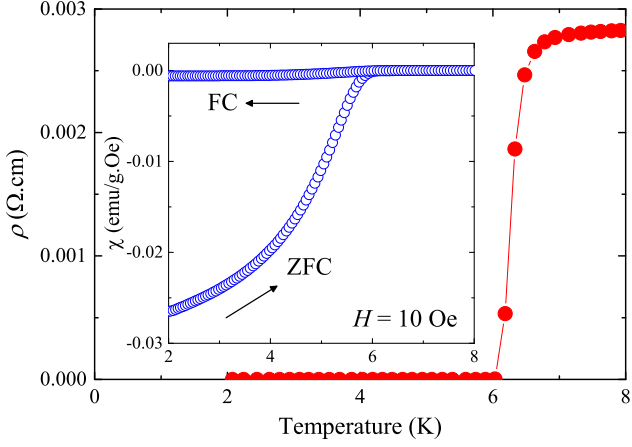}
\caption{(Color online) Resistivity vs. temperature curve for Nb$_2$PdS$_5$. The inset shows the temperature variation of the DC magnetic susceptibility curves of Nb$_2$PdS$_5$ in ZFC and FC modes in an applied field of 10 Oe.}
 \label{fig:res_mag}

\end{figure}

Figure~\ref{fig:res_mag} shows the temperature dependence of the resistivity of Nb$_2$PdS$_5$ under zero magnetic field. Zero-resistivity can be observed at 6.05 K. The inset shows the temperature variation of the DC magnetic susceptibility of Nb$_2$PdS$_5$ in zero-field-cooled (ZFC) and field-cooled (FC) modes in an applied field of 10 Oe. ZFC susceptibility shows a diamagnetic transition (onset) at around 6 K, a value very close to the zero-resistivity $T_{\rm c}$.

\begin{figure}[htb]
\includegraphics[width=1.0\linewidth]{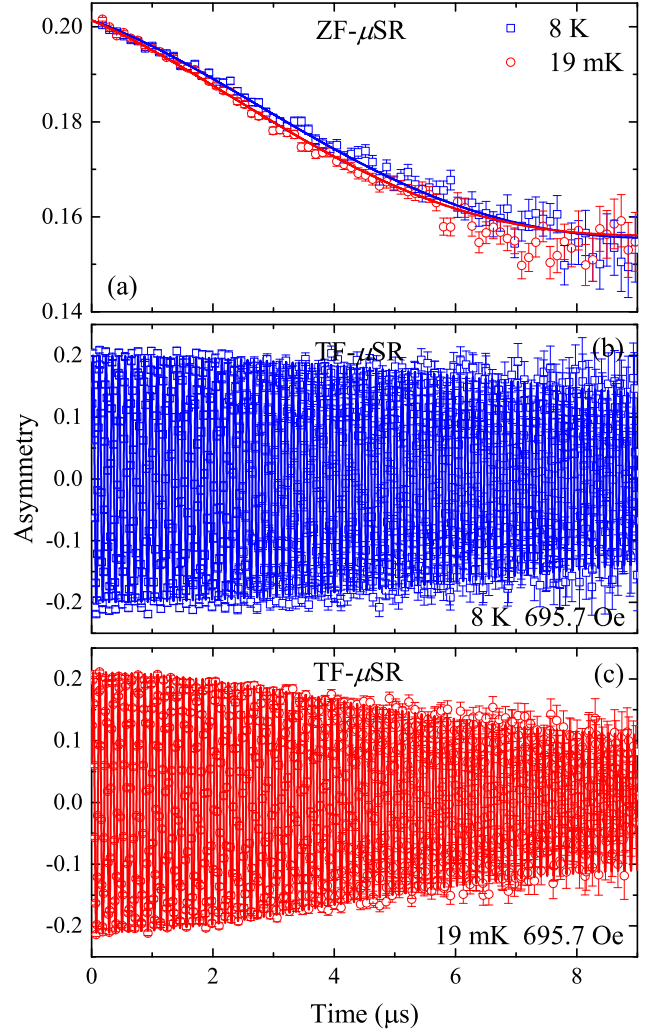}
\caption{(Color online) (a) ZF-$\mu$SR spectra of Nb$_2$PdS$_5$ taken at 8 K (above $T_{\rm c}$) and 19 mK (below $T_{\rm c}$). The solid lines are fits to the data using Eq.~\ref{eq:KT_ZFequation}. (b) and (c) show the TF-$\mu$SR time spectra, collected both above and below $T_{\rm c}$ in an applied field of 695.7 Oe. The solid lines are fits to the data using Eq.~\ref{eq:TFequation}.}
 \label{fig:musr_spectra}
\end{figure}

Figure~\ref{fig:musr_spectra}(a) compares the ZF-$\mu$SR signals collected above and below $T_{\rm c}$ which are practically identical. The ZF-$\mu$SR data can be well described using a combination of Gaussian and Lorentzian Kubo-Toyabe (KT) relaxation functions~\cite{Kubo},
\begin{multline}
A(t)= A(0)\left\{\frac{1}{3}+\frac{2}{3}\left(1-a^2t^2-\Lambda{t}\right){\rm exp}\left(-\frac{a^2t^2}{2}-\Lambda{t}\right)\right\} \\
+A_{\rm bg},
\label{eq:KT_ZFequation}
\end{multline}
where $A(0)$ is the initial asymmetry of the sample signal, $A_{\rm bg}$ is the background signal, $a$ and $\Lambda$ are the muon spin relaxation rates due to randomly oriented Nb and diluted $^{105}$Pd nuclear moments, respectively. The nuclear moment of S is negligible compare to the others. The KT relaxation function is of Gaussian type when the nuclear moments are densely populated in a lattice. A Lorentzian type of relaxation is expected for weak and diluted nuclear or electronic moments. For this material, Nb has a reasonably strong nuclear magnetic moment which is densely populated and therefore produce Gaussian type field distribution. On the other hand, Pd has several isotopes with only one of them carrying a nuclear moment. This isotope has a natural abundance of only 22\% and therefore most probably causes a field distribution similar to a Lorentzian. The fits to the ZF-$\mu$SR signals using Eq.~\ref{eq:KT_ZFequation} yield, $a=0.16(2)$~$\mu$s$^{-1}$, $\Lambda(19{\rm mK})=0.079(4)$~$\mu$s$^{-1}$, and $\Lambda(8{\rm K})=0.065(3)$~$\mu$s$^{-1}$. Comparing the two ZF-$\mu$SR signals, collected above and below $T_{\rm c}$, we see a marginally small increase of damping rate corresponds to a magnetic field width increase of 0.16 G at low temperature. One could argue that this additional contribution might be related to the spontaneous magnetic moments associated with a time-reversal-symmetry (TRS) breaking pairing state in Nb$_2$PdS$_5$~\cite{Luke1,Luke2,Reotier,Aoki,Maisuradze,Hillier,Hillier1,Singh,Biswas1}. However, it is not sizeable enough to make such a claim for Nb$_2$PdS$_5$. Further studies are in progress to explore the presence of TRS breaking spontaneous magnetic moments in this material. In the absence of TRS breaking fields, an alternative explanation for such a small field is the presence of a small amount of magnetic impurities in the sample.

Figure~\ref{fig:musr_spectra}(b) and (c) show the TF-$\mu$SR time spectra collected at 8~K and 19~mK in a transverse field of 695.7~Oe. The data obtained in the normal state show a very weak relaxation of the $\mu$SR time spectra. In this case the local field probed by the muons is essentially the applied field. The signal is only slightly damped by the nuclear moments contribution. By contrast, the data collected in the superconducting state shows a more pronounced damping due to the inhomogeneous field distribution generated by the formation of a vortex lattice in Nb$_2$PdS$_5$. The TF-$\mu$SR time spectra were analyzed using a Gaussian damped precession signal:
\begin{multline}
A^{TF}(t)=A(0)\exp\left(-\sigma^{2}t^{2}\right/2)\cos\left(\gamma_\mu \left\langle B\right\rangle t +\phi\right)\exp\left(-\alpha t\right) \\
+A_{\rm bg}(0)\cos\left(\gamma_\mu B_{\rm bg}t +\phi\right),
\label{eq:TFequation}
\end{multline}
where $A(0)$ and $A_{\rm bg}$(0) are the initial asymmetries of the sample and background signals, $\gamma_{\mu}/2\pi=13.55$~kHz/G is the muon gyromagnetic ratio~\cite{Sonier}, $\left\langle B\right\rangle$ and $B_{\rm bg}$ are the internal and background magnetic fields, $\phi$ is the initial phase of the muon precession signal, $\sigma$ is the Gaussian muon spin relaxation rate, and $\alpha$ is the exponential relaxation rate due to weak and diluted nuclear or electronic moments. The background signal mainly originates from muons hitting the silver sample holder, where the relaxation  rate is negligibly small. A global fit to the TF-$\mu$SR signals using Eq.~\ref{eq:TFequation} yield $\alpha=0.031(3)$~$\mu$s$^{-1}$. 

\begin{figure}[htb]
\includegraphics[width=1.0\linewidth]{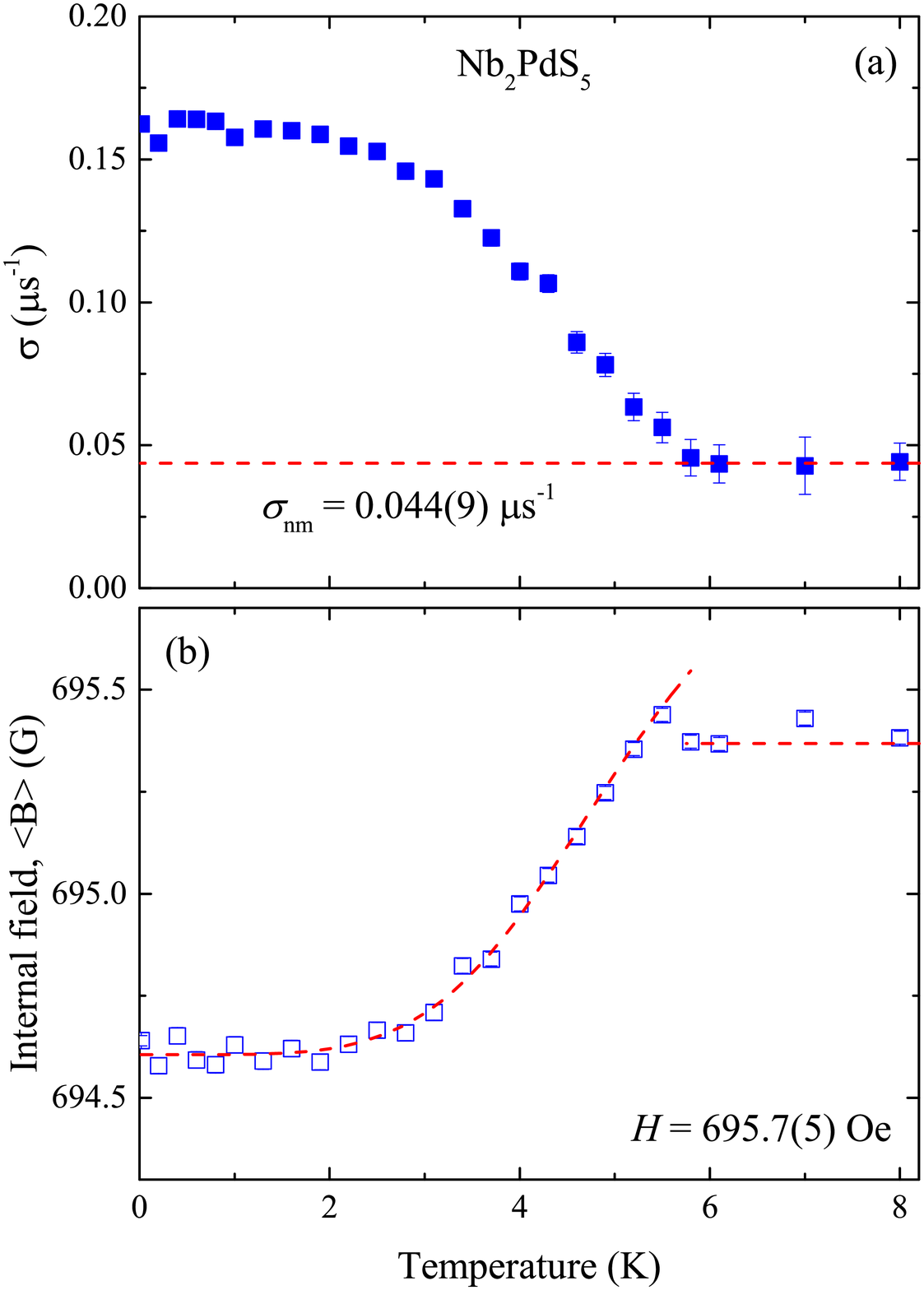}
\caption{(Color online) (a) Temperature dependence of the muon depolarization rate $\sigma$ of Nb$_2$PdS$_5$ collected in an applied magnetic field of 695.7 Oe. (b) Typical diamagnetic shift of the internal field experienced by the muons below $T_{\rm c}$. The dashed lines are guides to the eye only.}
 \label{fig:sigmaT}
\end{figure}

Figure~\ref{fig:sigmaT} (a) shows the temperature dependence of $\sigma$ of Nb$_2$PdS$_5$ for an applied field of 695.7 Oe. The temperature dependence of the internal magnetic field at the muon site with the expected diamagnetic shift below $T_{\rm c}$ is shown in Fig.~\ref{fig:sigmaT} (b). The dashed lines are guides to the eye only.

\begin{figure}[htb]
\includegraphics[width=1.0\linewidth]{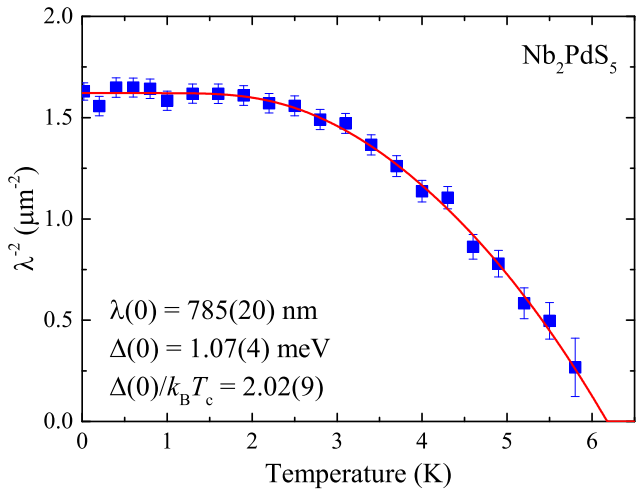}
\caption{(Color online) The temperature dependence of $\lambda^{-2}(T)$. The solid line is a fit to the data with the weak-coupling BCS model.}
 \label{fig:lambdaT}
\end{figure}

\begin{figure}[htb]
\includegraphics[width=1.0\linewidth]{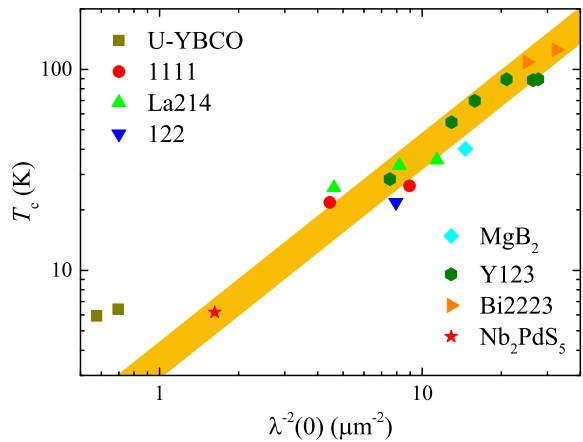}
\caption{(Color online) The Uemura plot for high $T_{\rm c}$ cuprates, MgB$_2$ and Fe-based superconductors with the data points taken from Ref.~\cite{Luetkens,Fletcher,Broun}. The red star represents the data for Nb$_2$PdS$_5$ obtained in this work.}
 \label{fig:Uemura_plot_v2}
\end{figure}

The superconducting contribution to $\sigma$ is obtained by subtracting the nuclear moment contribution (measured above $T_{\rm c}$) as ${\sigma_{sc}}^2=\sigma^2-{\sigma_{\rm nm}}^2$. In an isotropic type-II superconductor with a hexagonal Abrikosov vortex lattice described by Ginzburg-Landau theory, the magnetic penetration depth $\lambda$ is related to $\sigma
_{sc}$ by the Brandt equation~\cite{Brandt}:
\begin{equation}
\sigma_{sc}[\mu{\rm s}^{-1}]=4.854\times10^4\left(1-\frac{H}{H_{\rm c2}}\right)\left[1+1.21\left(1-\sqrt{\frac{H}{H_{\rm c2}}}\right)^3\right]\lambda^{-2}[{\rm nm}^{-2}],
 \label{eq:Brandt_equation}
\end{equation}
where $H$ and $H_{\rm c2}$ are the applied and upper critical field, respectively. The temperature dependence of $\lambda^{-2}(T)$ was calculated by using Eq.~\ref{eq:Brandt_equation} and $H_{\rm c2}(T)$ from Ref.~\cite{Niu}. Fig.~\ref{fig:lambdaT} shows the temperature dependence of $\lambda^{-2}$ which is proportional to the effective superfluid density, $\rho_s\propto\lambda^{-2}$.  As Fig.~\ref{fig:lambdaT} illustrates, $\rho_{\rm s}$ is very nearly constant below $T_{\rm c}/3\approx2$ K. This possibly suggests the absence of low-lying excitations and is indicative of a nodeless superconducting gap at the Fermi surface below $T_{\rm c}$  in Nb$_2$PdS$_5$. This is further verified by a good fit to the $\lambda^{-2}(T)$ data with a single-gap BCS $s$-wave model (solid line in Fig.~\ref{fig:lambdaT})~\cite{Tinkham,Prozorov}:
\begin{equation}
\frac{\lambda^{-2}(T)}{\lambda^{-2}(0)}= 1+2\int_{\Delta(T)}^{\infty}\left(\frac{\partial f}{\partial E}\right)\frac{E dE}{\sqrt{E^2-\Delta(T)^2}}.
 \label{eq:lambda}
\end{equation}
Here $\lambda^{-2}(0)$ is the  zero-temperature value of the magnetic penetration depth and $f=[1+\exp(E/k_BT)]^{-1}$ is the Fermi function. The BCS temperature dependence of the superconducting gap function is approximated as~\cite{Carrington}
\begin{equation}
\Delta(T)=\Delta(0)\tanh\left\{1.82\left[1.018\left(\frac{T_{\rm c}}{T}-1\right)\right]^{0.51}\right\}
 \label{eq:delta}
\end{equation}
where $\Delta(0)$ is the gap magnitude at zero temperature. The fit yields $T_c=6.2(2)$~K, $\lambda(0)=785(20)$~nm, and $\Delta(0)=1.07(4)$~meV. The gap to $T_{\rm c}$ ratio $\Delta(0)/k_{\rm B}T_{\rm c}=2.02(9)$ is higher than the BCS value of 1.76, suggesting that Nb$_2$PdS$_5$ is a moderately strong-coupling superconductor. It is noteworthy to mention that, though the BCS model is appropriate for a weak-coupling superconductor, standard BCS theory has been widely used in the past to determine electronic parameters even for systems which do not fulfill all the strict criteria of a BCS superconductor. In exotic superconductors, such as cuprates and Fe-based systems, the value and symmetry of the gap obtained from $\mu$SR using the BCS expression are in many cases in good agreement with values obtained by other spectroscopic methods such as ARPES~\cite{Khasanov}. We add the value of $\lambda(0)$ for Nb$_2$PdS$_5$ as a red star to the Uemura plot, shown in Fig.~\ref{fig:Uemura_plot_v2}, which shows the scaling relation between $\lambda^{-2}(0)$ and $T_c$ for unconventional superconductors~\cite{Uemura}. The value of $\lambda^{-2}(0)$ places Nb$_2$PdS$_5$ well inside the broad line for cuprates and iron-based superconductors on the Uemura plot.

Using $H_{\rm c2}(0)=176(2)$~kOe (estimated by fitting the $H_{\rm c2}(T)$ data of Ref.~\cite{Niu} with a Ginzburg-Landau (GL) model~\cite{Tinkham}) and its relation with the coherence length $\xi$, $\left(\mu_{0}H_{\rm c2}=\frac{\phi_0}{2\pi\xi^2}\right)$, we calculate $\xi=4.3(1)$~nm at 0~K. Using the values of $\lambda$ and $\xi$, we calculate the GL parameter, $\kappa=\frac{\lambda}{\xi} \cong$ 183. The large value of $\kappa$ indicates that Nb$_2$PdS$_5$ is an extreme type-II superconductor. For completeness, by combining the value of $\xi$ and our measured value of $\lambda$, we calculate the value of lower critical field, $H_{\rm c1}$ using the expression~\cite{Brandt}:
\begin{equation}
\mu_{0}H_{\rm c1}=\frac{\phi_0}{4\rm \pi\lambda^{2}}\left(\ln \frac{\lambda}{\xi}+0.5\right)
 \label{eq:hc1}
\end{equation}
and obtain $H_{\rm c1}(0)=14(1)$~Oe.


In conclusion, $\mu$SR measurements have been performed on superconducting Nb$_2$PdS$_5$. ZF-$\mu$SR results do not find evidence of any sizeable spontaneous magnetization in Nb$_2$PdS$_5$. TF-$\mu$SR results show that the superfluid density becomes practically temperature independent below $T_{\rm c}/3\approx2$ K, which is consistent with a fully gapped superconducting state. $\rho_s(T)$ can be well described within the single-gap BCS $s$-wave scenario with $\Delta(0)=1.07(4)$~meV. The magnetic penetration depth was estimated to $\lambda(0)=785(20)$~nm, which is significantly large but consistent with the $\lambda(0)$ values of other low-dimensional superconductors~\cite{Pratt,Harshman,Le,Greer,Luke,Scheidt}. For layered superconductors, such as cuprates~\cite{Baker} and iron chalcogenides~\cite{Biswas}, it has been observed in the past that the penetration depth increases with increasing layer separation, i.e. when the superfluid density goes from 3D to 2D nature. The same is true when superfluid density becomes Q1D in nature and explains the finding of such a large  $\lambda(0)$ in this system. There is no signature of multiple gaps, as it may be expected from the multiband structure of this material possibly indicating that the gaps have a similar magnitude. The value of the gap to $T_{\rm c}$ ratio, 2.02(9) is higher than the BCS value of 1.76 and suggests that Nb$_2$PdS$_5$ is a moderately strong-coupling superconductor. This is probably partially responsible for such a large $H_{\rm c2}$ along the $b$-axis, similar to heavy-fermion superconductors, such as UBe$_{13}$~\cite{Thomas}. The value of $\lambda^{-2}(0)$ places Nb$_2$PdS$_5$ well inside the broad line for unconventional superconductors in a Uemura plot. The results presented here will provide a reference point for further $\mu$SR studies on single crystals at high magnetic field, applied both parallel and perpendicular to the principle $b$-axis.

The $\mu$SR experiments were performed at the Swiss Muon Source (S$\mu$S), Paul Scherrer Institute (PSI, Switzerland). X.Xu would like to acknowledge the financial support from the National Key Basic Research Program of China (Grant No. 2014CB648400)  and from NSFC (Grant No. 11474080).

\end{document}